\documentclass[11pt,notitlepage,nofootinbib]{revtex4-1}

\usepackage[T1]{fontenc}

\usepackage{epsf}
\usepackage{latexsym}
\usepackage{amssymb}
\usepackage{amsmath}
\usepackage{bm}
\usepackage[dvips]{graphicx}
\usepackage{array}
\usepackage{xcolor}

\renewcommand{\theequation}{\arabic{section}.\arabic{equation}}

\date{\today}
\def\d3{^{(3)}\nabla}

\begin{document}

\title{Penrose limits of inhomogeneous space-times, their diagonalizability and twistors}

\author{Kerstin E. Kunze}

\email{kkunze@usal.es}

\affiliation{Departamento de F\'\i sica Fundamental, Universidad de Salamanca,
 Plaza de la Merced s/n, 37008 Salamanca, Spain}

\begin{abstract}
Penrose limits are considered in space-times admitting two abelian, space-like Killing vectors
in vacuum as well as in the presence of an electromagnetic field.
This type of space-times describe inhomogeneous cosmologies as well as colliding plane gravitational and electromagnetic waves.  
Following the work of Tod \cite{Tod2020} the conditions for diagonal Penrose limits are investigated in these backgrounds. 
The twistor equation is considered in these space-times and solutions given in the Penrose limit.

\end{abstract}

\maketitle


\section{Introduction}
\label{s0}
\setcounter{equation}{0}

Plane wave solutions play an important role in classical as well as quantum field theories. 
In string theory they are also known to be examples of exact classical string vacua \cite{Amati,Horowitz}. 
Penrose \cite{penrose} showed that along a segment of a null geodesic without conjugate points 
any space-time has a plane wave limit.
The Penrose limit and its generalization to $D$ dimensional space-times  \cite{Gueven} is of particular interest in string-/M theory (cf. e.g. \cite{Skrzypek:2021,Blau:2002mw})

Recently Tod \cite{Tod2020}  demonstrated the condition on the space-time so that {\it all} Penrose limits are diagonal.
Moreover, an elegant and computational efficient way of calculating the Penrose limit has been presented there using 2-spinor calculus \cite{Tod2020}. Different from the original formulation of Penrose \cite{penrose} it does not involve an explicit coordinate transformation adapted to the null geodesic along which the Penrose limit is taken.

Here the formulation of Tod \cite{Tod2020} is applied to  determine the Penrose limits of inhomogeneous space-times admitting two space-like commuting Killing vectors generating an abelian group $G_2$.
This type of metrics allows to describe different types of backgrounds such as cosmological backgrounds
and colliding plane wave space-times. Moreover, some of  the spatially homogeneous backgrounds with 3 space-like Killing vectors, namely,  models of Bianchi type I-VII as well as the locally rotationally symmetric (LRS) VIII and LRS IX admit two dimensional  abelian subgroups
 (cf. \cite{stepExSol}).
Radial Penrose limits and abelian as well as non-abelian T-duality transformations of low energy string backgrounds have been considered in this type of backgrounds in \cite{kk03}.

An interesting aspect of the resulting plane wave space-times in the Penrose limit is that these backgrounds permit the existence of twistors \cite{PR2,Tod2020,Lewandowski}. Twistors were introduced to provide a fundamental  description of  space-time structure and physical concepts. Penrose proposed to consider the 2-spinors as more fundamental than  space-time points (e.g. \cite{PR2}).
Therefore it is natural to expect that twistors play an important role in the quantization of gravity.
The relation of the twistor equation to massless fields and representation of solutions in terms of 
contour integrals of
holomorphic functions has led to an important advancement in the mathematical aspects of solutions of differential equations by Penrose transforms (e.g., \cite{BastonEastwood}). Global solutions of the twistor equation in a curved background are severely restricted by a consistency condition. This has led to additional concepts of local and asymptotic twistors in asymptotically flat space-times. However, there do exist global twistors in plane wave backgrounds. In this sense the importance of the resulting plane wave space-times in the Penrose limit is  similar to the case of exact solutions of string theory in plane wave backgrounds which in part motivated the strong interest in the Penrose limit over recent years.

The plan of the paper is as follows.
In section \ref{sec1} the Penrose limit procedure and all the relevant quantities in the Newman-Penrose formalism as well as in the  2-spinor formulation are presented for  $G_2$ metrics.
In section \ref{sec1a} the electromagnetic field  2-spinor formulation for the $G_2$ metrics is given together with Einstein's equations.
In section \ref{sec2} the resulting wave profiles in the Brinkmann form of plane wave space-times are given. In section \ref{sec3} the question of diagonalizability will be discussed. Radial Penrose limits are considered as an example in section \ref{sec4}. The twistor equation is discussed in section \ref{sec5} and solutions are given in the Penrose limit.
Finally, in section \ref{sec6}  conclusions are presented.


\section{Penrose limits of $G_2$ space-times}
\label{sec1}
\setcounter{equation}{0}
Metrics admitting two abelian space-like Killing vectors  are described by the line element (e.g. \cite{griff})
\begin{eqnarray}
ds^2=2e^{M}du dv-\frac{e^{-U}}{Z+\bar{Z}}\left(dx+i Z dy\right)\left(dx-i\bar{Z}dy\right)
\label{metg2}
\end{eqnarray}
where $M=M(u,v)$ and $U=U(u,v)$ are real functions of the null coordinates $u$ and $v$. 
 $Z=Z(u,v)$ is a complex function of $u$ and $v$.
Colliding plane wave space-times can be separated in four different regions. Two of these describe the two incoming plane waves for which all metric functions have as argument either the null coordinate $u$ or the null  coordinate $v$, respectively (e.g. \cite{griff}). The interaction region of these plane waves constitutes the third region in which all metric functions depend on both null coordinates in general. The fourth region describes the background space-time on which the waves propagate, most commonly taken to be flat. The Penrose limits calculated of the interaction region might have interesting relations to the incoming, initial plane waves.

When considering cosmological space-times it is useful to introduce a timelike coordinate $t$ and a space-like coordinate $z$ by defining $t=u-v$ and $z=u+v$. The explicit forms of the metric functions $M$, $U$ and $Z$ for the case of the spatially homogenous models of Bianchi type which have 3 Killing vectors but admit two dimensional abelian subgroups $G_2$ have been found explicitly in \cite{kk03}.
 
The Newman-Penrose formalism allows to efficiently calculate components of curvature and energy momentum tensor as well as covariant derivatives and Einstein's equations by introducing an orthonormal  null tetrad with two real null vectors $l^{\mu}$ and $n^{\mu}$, respectively, and two complex null vectors which are complex conjugates of each other $m^{\mu}$ and $\bar{m}^{\mu}$, respectively \cite{chandra, stepExSol, PR1}. These are given by
\begin{eqnarray}
l_{\mu}n^{\mu}=1\;\;\;  m_{\mu}\bar{m}^{\mu}=-1
\end{eqnarray}
and all other combinations are zero. In terms of the null tetrad vectors the space-time metric is given by 
$g_{\mu\nu}=l_{\mu}n_{\nu}+n_{\mu}l_{\nu}-m_{\mu}\bar{m}_{\nu}-\bar{m}_{\mu}m_{\nu}$.
The spin  coefficients and curvature components for the $G_2$-metric (\ref{metg2}) are given in appendix \ref{appA}.
Defining a normalized spin-frame $o^A$, $\iota^A$, with $A=0,1$ satisfying \cite{PR1} 
\begin{eqnarray}
o_A\iota^A=1\hspace{1cm}\iota_A o^A=-1 \hspace{2.5cm}o_A o^A=0=\iota_A \iota^A,
\end{eqnarray}
then the null tetrad vectors are determined by 
\begin{eqnarray}
l^{\alpha}=o^Ao^{A'}\hspace{1cm}n^{\alpha}=\iota^A\iota^{A'}
\hspace{1cm}m^{\alpha}=o^A\iota^{A'}
\hspace{1cm}\bar{m}^{\alpha}=\iota^A o^{A'}.
\end{eqnarray}

The Penrose limit yields  a plane wave space-time. This  admits 5 Killing vectors. The only non-vanishing Weyl scalar is $\Psi_4=\Psi$ and $\Phi_{22}=\Phi$ is the only non vanishing scalar determining the components of the Ricci tensor  (cf. e.g. \cite{griff,stepExSol,Tod2020}). Equivalently the Weyl spinor and Ricci spinor are given by,  respectively, \cite{Tod2020}
\begin{eqnarray}
\psi_{ABCD}=\Psi o_A o_B o_C o_D
\hspace{3cm}
\phi_{ABA'B'}=\Phi o_A o_B o_{A'}o_{B'}.
\end{eqnarray}

Following \cite{Tod2020} the Penrose limit of  a space-time $M$ is obtained by choosing any null geodesic $\Gamma$
in $M$ and considering a spinor field $\alpha^A$ parallely propagated tangent to $\Gamma$ and defining an affine parameter $s$ (upto an additive constant) by 
$\alpha^A\bar{\alpha}^{A'}\nabla_{AA'}s=1$. The plane wave in the Penrose limit is then determined by
\begin{eqnarray}
\Psi(s)=\psi_{ABCD}\alpha^A\alpha^B\alpha^C\alpha^D
\hspace{3cm}
\Phi(s)=\phi_{ABA'B'}\alpha^A\alpha^B\bar{\alpha}^{A'}\bar{\alpha}^{B'}.
\label{ricciWeylPL}
\end{eqnarray}

 Assuming the spinor field $\alpha^{A}$ to be of the form 
 \begin{eqnarray}
 \alpha^A=A(s)o^A+B(s) \iota^A
 \label{alph}
 \end{eqnarray}
 the condition for its parallel transport along the null geodesic $\Gamma$,
 \begin{eqnarray}
 \alpha^A\bar{\alpha}^{A'}\nabla_{AA'}\alpha_B=0
 \end{eqnarray}
 yields the evolution equations for the two complex functions $A(s)$ and $B(s)$
 \begin{eqnarray}
 \frac{dA}{ds}&=&-|A|^2\left(A\epsilon+B\alpha-B\tau'\right)+|B|^2\left(A\rho'+B\kappa'-A\gamma\right)
 -A^2\bar{B}\beta+B^2\bar{A}\sigma'\\
 \frac{dB}{ds}&=&|A|^2\left(A\kappa+B\epsilon+B\rho\right)-|B|^2\left(-A\beta-A\tau-B\gamma\right)
 +A^2\bar{B}\sigma+B^2\bar{A}\alpha
 \label{Bevol}
 \end{eqnarray}
 using that $\{o^A, \iota^A\}$ is a normalized spinor basis.
 For the $G_2$ metric equation (\ref{metg2}) the spin coefficients $\tau$, $\tau'$, $ \alpha$, $\beta$, $\kappa$ and $\kappa'$ vanish (cf. appendix \ref{appA}, equation (\ref{spincoef})).
 
 The two Killing vectors $\partial_x$ and $\partial_y$ admitted by  the metric (\ref{metg2})  expressed in terms of the null tetrad are given by
 \begin{eqnarray}
 ^{(1)}K^{\mu}&=&\frac{e^{-\frac{U}{2}}}{(Z+\bar{Z})^{\frac{1}{2}}}m^{\mu}+\frac{e^{-\frac{U}{2}}}
 {(Z+\bar{Z})^{\frac{1}{2}}}\bar{m}^{\mu}\\
 ^{(2)}K^{\mu}&=&i\frac{e^{-\frac{U}{2}}}{(Z+\bar{Z})^{\frac{1}{2}}}Zm^{\mu}-i\frac{e^{-\frac{U}{2}}}
 {(Z+\bar{Z})^{\frac{1}{2}}}\bar{Z}\bar{m}^{\mu}.
 \end{eqnarray}
 Therefore there are two constants of motion $^{(i)}E=\, ^{(i)}K_{\mu}V^{\mu}$ where the tangent world vector of the null geodesic is determined by $V^{\mu}=\alpha^A\bar{\alpha}^{A'}$ thus
 \begin{eqnarray}
 V^{\mu}=A\bar{A}\, l^{\mu}+B\bar{B} n^{\mu}+A\bar{B}m^{\mu}+\bar{A}B\bar{m}^{\mu}.
 \label{nullgeo}
 \end{eqnarray}
 This yields to 
 \begin{eqnarray}
 ^{(1)}E&=&-\frac{e^{-\frac{U}{2}}}{(Z+\bar{Z})^{\frac{1}{2}}}(\bar{A}B+A\bar{B})\\
 ^{(2)}E&=&-i\frac{e^{-\frac{U}{2}}}{(Z+\bar{Z})^{\frac{1}{2}}}(\bar{A}BZ-A\bar{B}\bar{Z}),
 \end{eqnarray}
 implying that
 \begin{eqnarray}
 \bar{A}B=\frac{e^{\frac{U}{2}}}{(Z+\bar{Z})^{\frac{1}{2}}}\left(-\;^{(1)}E\bar{Z}+i\, ^{(2)}E\right).
 \label{conjAB}
 \end{eqnarray}
 This allows to obtain expressions for the product of the modulus of $A$ and $B$ as well as the differences in their phases, namely,
 \begin{eqnarray}
 |A||B|&=&\frac{e^{\frac{U}{2}}}{(Z+\bar{Z})^{\frac{1}{2}}}
 \left[(\;^{(1)}E)^2Z\bar{Z}+i(\bar{Z}-Z)\;^{(1)}E\;^{(2)}E+(\;^{(2)}E)^2\right]^{\frac{1}{2}}
 \label{abs(ab)}
 \\
e^{\pm 2i(\varphi_B-\varphi_A)}&=&\left(\frac{^{(1)}E\bar{Z}-i\;^{(2)}E}{^{(1)}EZ+i\;^{(2)}E}\right)^{\pm 1}
\label{phiB-A}
 \end{eqnarray}
 with $X=|X|e^{i\varphi_X}$, $X=A,B$.
 Finally, using the spin coefficients as given in equation (\ref{spincoef}) this leads to 
 \begin{eqnarray}
 \frac{d|A|^2}{ds}&=&\frac{1}{2}e^{\frac{M}{2}}(\partial_v M)|A|^4+\frac{1}{2}
 \frac{e^{\frac{M}{2}+U}}{Z+\bar{Z}}
 \left[\partial_u(2U-M)\left[(\;^{(1)}E)^2Z\bar{Z}+i(\bar{Z}-Z)\;^{(1)}E\;^{(2)}E+(^{(2)}E)^2\right]
 \right.
 \nonumber\\
 &+&\left.
 \frac{2}{Z+\bar{Z}}\partial_u\left[\frac{1}{3}(\;^{(1)}E)^2(Z^3+\bar{Z}^3)-i\;^{(1)}E\;^{(2)}E(Z^2-\bar{Z}^2)-(\;^{(2)}E)^2(Z+\bar{Z})\right]\right]\\
 \frac{d|B|^2}{ds}&=&\frac{1}{2}e^{\frac{M}{2}}(\partial_u M)|B|^4+\frac{1}{2}
 \frac{e^{\frac{M}{2}+U}}{Z+\bar{Z}}
 \left[\partial_v(2U-M)\left[(\;^{(1)}E)^2Z\bar{Z}+i(\bar{Z}-Z)\;^{(1)}E\;^{(2)}E+(^{(2)}E)^2\right]
 \right.
 \nonumber\\
 &+&\left.
 \frac{2}{Z+\bar{Z}}\partial_v\left[\frac{1}{3}(\;^{(1)}E)^2(Z^3+\bar{Z}^3)+i\;^{(1)}E\;^{(2)}E(Z^2-\bar{Z}^2)-(\;^{(2)}E)^2(Z+\bar{Z})\right]\right]
 \end{eqnarray}
 Moreover, the condition $s$ to be an affine parameter leads to 
 \begin{eqnarray}
 \frac{du}{ds}=|B|^2e^{\frac{M}{2}}& \hspace{1cm}&
  \frac{dv}{ds}=|A|^2e^{\frac{M}{2}}\nonumber\\
  \frac{dx}{ds}=\frac{e^{\frac{U}{2}}}{(Z+\bar{Z})^{\frac{1}{2}}}(A\bar{B}\bar{Z}+\bar{A}BZ)
  &\hspace{1cm}&
  \frac{dy}{ds}=-i\frac{e^{\frac{U}{2}}}{(Z+\bar{Z})^{\frac{1}{2}}}(A\bar{B}-\bar{A}B).
 \label{affinepar}
 \end{eqnarray}
 Taking into account equation (\ref{abs(ab)}) the last two equations can be re-written as
 \begin{eqnarray}
 \frac{dx}{ds}&=&-\frac{2e^U}{Z+\bar{Z}}\left(Z\bar{Z}\;^{(1)}E-\frac{i}{2}(Z-\bar{Z})\;^{(2)}E\right)\\
 \frac{dy}{ds}&=&-\frac{2e^U}{Z+\bar{Z}}\left(-\frac{i}{2}(Z-\bar{Z})\;^{(1)}E+^{(2)}E\right).
 \end{eqnarray}
Thus  null geodesics with $\dot{x}=0$, $\dot{y}\neq 0$ are possible for $^{(1)}E=0$,
 $^{(2)}E\neq 0$ in space-times with $Z$ real. Interchanging the constants with the same condition on $Z$ allows for the case   $\dot{x}\neq 0$, $\dot{y}=0$. In the case $\dot{x}=0=\dot{y}$ both constants have to vanish. Moreover, together with equation (\ref{conjAB}) this implies that one of the functions $A$ or $B$ has to vanish, making the affine parameter $s$ a function only of either one of  the null coordinates $u$ or $v$ of the background space-time. This describes radial Penrose limits in which the Penrose limit is considered along radial geodesics (e.g. \cite{kk03}). Radial Penrose limits within the spinor formulation of Tod \cite{Tod2020} is considered below in section \ref{sec4}.

 The only non-vanishing  Ricci spinor and Weyl scalar in the Penrose limit (cf. equation (\ref{ricciWeylPL})) are given by
 \begin{eqnarray}
 \Phi(s)&=&|A|^4\Phi_{00}+(A\bar{B})^2\Phi_{02}+(\bar{A}B)^2\Phi_{20}-4|A|^2|B|^2\Phi_{11}+|B|^4\Phi_{22}
 \label{phi_s}\\
 \Psi(s)&=&A^4\Psi_0+6A^2B^2\Psi_2+B^4\Psi_4
 \label{psi_s}
 \end{eqnarray}
 where $\Phi_{ab}$ and $\Psi_i$ are the tetrad components of the Ricci tensor and the Weyl scalars
 for the $G_2$ metric (\ref{metg2}) as given in equations (\ref{phi00})-(\ref{phi22}) and
 (\ref{psi0})-(\ref{psi4}), respectively.


\section{The electromagnetic field in $G_2$ space-times and its Penrose limit}
\label{sec1a}
\setcounter{equation}{0} 

The Maxwell tensor is determined by the electromagnetic spinor $\varphi_{AB}$ by \cite{PR1}
\begin{eqnarray}
F_{\alpha\beta}=F_{AA'BB'}=\varphi_{AB}\epsilon_{A'B'}+\epsilon_{AB}{\overline{\varphi}}_{A'B'}
\end{eqnarray}
with $\varphi_{AB}=\varphi_{(AB)}=\frac{1}{2}F_{ABC'}^{\hspace{0.7cm} C'}$ with the correspondence between space-time and spinor components given by $\varphi_0=\varphi_{00}$, $\varphi_1=\varphi_{01}$, $\varphi_2=\varphi_{11}$ and its complex conjugate.
The free space Maxwell's equations can be expressed in terms of a zero rest-mass field equation \cite{PR1}  which in the  compacted spin-coefficient form (GHP formalism) is efficiently written as
\begin{eqnarray}
\textup{\th}\varphi_1-\textup{\dh}'\varphi_0&=&-\tau'\varphi_0+2\rho\varphi_1-\kappa\varphi_2\\
\textup{\th}'\varphi_1-\textup{\dh}'\varphi_2&=&-\tau\varphi_2+2\rho'\varphi_1-\kappa'\varphi_2\\
\textup{\th}\varphi_2-\textup{\dh}'\varphi_1&=&\sigma'\varphi_0-2\tau'\varphi_1+\rho\varphi_2\\
\textup{\th}'\varphi_0-\textup{\dh}\varphi_1&=&\sigma\varphi_0-2\tau\varphi_1+\rho'\varphi_0
\end{eqnarray}
Writing these equations explicitly for the $G_2$ metric (\ref{metg2}) and its spin coefficients (cf. Appendix \ref{appA}) and taking into account that all metric functions only depend on the null variables $u$ and $v$ the first two equations yield to  
\begin{eqnarray}
\varphi_1(u,v)=c_{\varphi_1}e^{U}
\end{eqnarray}
where $c_{\varphi_1}$ is a constant. 
Using the notation $X_{,m}\equiv\frac{\partial X}{\partial m}$ and 
$X_{,mn} \equiv\frac{\partial^2 X}{\partial m\partial n}$ for $m,n$ denoting the null variables $u$ and $v$, the last two equations yield to (cf. also \cite{griff})
\begin{eqnarray}
H_{,uv}+\frac{Z_{,u}}{Z+\bar{Z}}H_{,v}+\frac{\bar{Z}_{,v}}{Z+\bar{Z}}H_{,u}=0
\label{Huv}
\end{eqnarray}
whose solutions determine the remaining components of the electromagnetic spinor as
\begin{eqnarray}
\varphi_0&=&-e^{\frac{1}{2}(M+U)}\sqrt{Z+\bar{Z}}H_{,v}\\
\varphi_2&=&e^{\frac{1}{2}(M+U)}\sqrt{Z+\bar{Z}}H_{,u}.
\end{eqnarray}
Indices $u$ and $v$ denote the corresponding partial derivatives.

Einstein's equations imply that the components of the Ricci tensor
in terms of the energy-momentum tensor $T_{ab}$ are  given by \cite{PR1}
\begin{eqnarray}
\Phi_{\alpha\beta}=4\pi G_N(T_{\alpha\beta}-\frac{1}{4}T_{\mu}^{\mu}g_{\alpha\beta})
\label{Phi}
\end{eqnarray}
with $G_N$ Newton's constant of gravitation. Together with the energy momentum tensor 
of the electromagnetic field
$T_{\alpha\beta}=\frac{1}{2\pi}\varphi_{AB}\overline{\varphi}_{A'B'}$.
Using equations (\ref{phi00}) to (\ref{phi22})  together with equation (\ref{Phi}) yield to
\begin{eqnarray}
(Z+\bar{Z})(2Z_{,uv}-Z_{,u}U_{,v}-Z_{,v}U_{,u})-4Z_{,u}Z_{,v}-4G_Ne^{U}(Z+\bar{Z})^3H_{,u}
\overline{H}_{,v}&=&0
\label{max-e1}\\
e^{-2U}(U_{,uv}-U_{,u}U_{,v})+16G_N|c_{\varphi_1}|^2e^{-M}&=&0\\
2U_{,uu}-U^2_{,u}+2M_{,u}U_{,u}-4\frac{Z_{,u}\bar{Z}_{,u}}{(Z+\bar{Z})^2}-8G_Ne^{U}(Z+\bar{Z})H_{,u}\overline{H}_{,u}&=&0\\
2U_{,vv}-U^2_{,v}+2M_{,v}U_{,v}-4\frac{Z_{,v}\bar{Z}_{,v}}{(Z+\bar{Z})^2}-8G_Ne^{U}(Z+\bar{Z})H_{,v}\overline{H}_{,v}&=&0
\label{max-e4}
\end{eqnarray}
which form a complete a set of equations together with equation (\ref{Huv}).

\subsection{The electromagnetic field in the Penrose limit}

In the pp wave space-time of the  Penrose limit in the direction of the flagpole of the spinor $\alpha^{A}$ the only non-vanishing component of the electromagnetic spinor is given by $\varphi_2(s)=\varphi_{11}
=\varphi_{AB}\alpha^A\alpha^B$ and its complex conjugate. In terms of the components of the electromagnetic spinor of the original $G_2$ space-time this is found to be
\begin{eqnarray}
\varphi(s)=A^2(s)\varphi_{00}+2A(s)B(s)\varphi_{01}+B^2(s)\varphi_{11}.
\end{eqnarray}


\section{Wave profiles of the plane waves in the Penrose limit}
\label{sec2}
\setcounter{equation}{0}

The wave profiles of a plane wave space-times are obtained from the Brinkmann form (cf. e.g. \cite{Tod2020},\cite{griff},\cite{stepExSol})
\begin{eqnarray}
ds^2=2drds+(h_{11}X^2+2h_{12}XY+h_{22}Y^2)ds^2-dX^2-dY^2,
\end{eqnarray}
where
\begin{eqnarray}
h_{11}&=&\Phi(s)+\frac{1}{2}(\Psi(s)+\overline{\Psi}(s))\\
h_{12}&=&-\frac{i}{2}(\Psi(s)-\overline{\Psi}(s))
\label{h12}\\
h_{22}&=&\Phi(s)-\frac{1}{2}(\Psi(s)+\overline{\Psi}(s))
\end{eqnarray}
with the Ricci tensor component $\Phi(s)$ (\ref{phi_s}) and the Weyl scalar $\Psi(s)$ (\ref{psi_s}).
If $\Psi(s)$ is real then
$h_{12}\equiv 0$ and the corresponding $G_2$ metric is diagonal (cf. \cite{griff}).

\section{Diagonal Penrose limits}
\label{sec3}
\setcounter{equation}{0}

Tod \cite{Tod2020} showed that the plane wave space-time in the Penrose limit is diagonizable if
\begin{eqnarray}
\Sigma(s)\equiv i(\overline{\Psi}(s)\dot{\Psi}(s)-\Psi(s)\dot{\overline{\Psi}}(s))=0,
\label{diag}
\end{eqnarray}
where a dot denotes the derivative w.r.t. the affine parameter $s$.
For the $G_2$ metric under consideration (cf. equation (\ref{psi_s}))
\begin{eqnarray}
\Psi(s)=\sum_{N=0,2,4}\mu_N\Psi_N,
\hspace{0.5cm}{\rm with}\hspace{0.5cm} \mu_0=A^4(s),\hspace{1cm}
\mu_2=6A^2B^2,\hspace{1cm}
\mu_4=B^4(s).
\end{eqnarray}
This yields to
\begin{eqnarray}
-i\Sigma(s)&=&\sum_{N,M=0,2,4}\left[T_2(\mu_N,\mu_M)T_3(\Psi_N,\Psi_M)+T_1(\Psi_N,\Psi_M)T_4(\mu_N,\mu_M)\right.\nonumber\\
&&\left. \hspace{1.3cm}
+T_1(\mu_N,\mu_M)T_4(\overline{\Psi}_N,\overline{\Psi}_M)+T_2(\overline{\Psi}_N,\overline{\Psi}_M)T_3(\mu_N,\mu_M)\right]
\end{eqnarray}
where with $X_N=|X_N|e^{i\varphi_{X_N}}$
\begin{eqnarray}
T_1(X_N,X_M)&=&|X_N||X_M|(e^{-i(\varphi_{X_N}-\varphi_{X_M})}-e^{i(\varphi_{X_N}-\varphi_{X_M})})
\nonumber\\
T_2(X_N,X_M)&=&|X_N|\dot{|X_M|}(e^{-i(\varphi_{X_N}-\varphi_{X_M})}-e^{i(\varphi_{X_N}-\varphi_{X_M})})
\nonumber\\
&&+i\dot{\varphi}_{X_M}|X_N||X_M|(e^{-i(\varphi_{X_N}-\varphi_{X_M})}+e^{i(\varphi_{X_N}-\varphi_{X_M})})
\nonumber\\
T_3(X_N,X_M)&=&|X_N||X_M|e^{-i(\varphi_{X_N}-\varphi_{X_M})}
\nonumber\\
T_4(X_N,X_M)&=&|X_N|\dot{|X_M|}e^{i(\varphi_{X_N}-\varphi_{X_M})}
-i\dot{\varphi}_{X_M}|X_N||X_M|e^{i(\varphi_{X_N}-\varphi_{X_M})}.
\end{eqnarray}
It can be seen that the diagonalizability condition (\ref{diag}) is satisfied for $\varphi_{\mu_N}=\varphi_{\mu}=const.$
and $\varphi_{\Psi_N}=\varphi_{\psi}=const.$.
From equation (\ref{phiB-A}) it follows that in this case the imaginary part of $Z$ is constant, namely, $\frac{i}{2}(Z-\bar{Z})=\frac{^{(2)}E}{^{(1)}E}$.
Thus the Weyl scalars $\Psi_0$, $\Psi_2$ and $\Psi_4$ are real functions and $\varphi_{\psi}\equiv 0$.
This still leaves the possibility that the constant phase $\varphi_{\mu}\neq 0$ and thus the Weyl scalar $\Psi(s)$ of the plane wave space time in the Penrose limit is complex. Formally this induces a non-diagonal wave amplitude $h_{12}$
(cf. equation (\ref{h12})).
However, the constant phase $4\varphi_{\mu}$ of $\Psi(s)$ can be made to vanish by a rotation of the null tetrad or equivalently the spinor dyad. In the former case following \cite{chandra} a rotation of class III leaves the directions of the null tetrad vectors ${\bf l}$ and ${\bf n}$ unchanged but rotates 
${\bf m}$ and ${\bf \overline{m}}$ by an angle $\theta$ in the $(\bf{m},\bf{\bar{m}})$ plane.
Only considering this rotation leads to a transformation  of the Weyl scalars $\Psi_j\rightarrow e^{i(2-j)\theta}\Psi_j$, $j=0,..,4$. Thus $\Psi(s)$ will be transformed to $e^{-2i\theta}\Psi(s)$ and choosing $\theta=2\varphi_A$ leads to a real Weyl scalar. Thus $h_{12}$ (\ref{h12}) vanishes.
In the Penrose limit the spin frame will be chosen to be determined by the normalized spinor dyad 
$\{{\bf\beta},{\bf\alpha}\}$ with ${\bf\alpha}$ given by equation (\ref{alph}) then ${\bf\beta}$ is found to be
\begin{eqnarray}
\beta^A=Fo^A+G\iota^A
\end{eqnarray}
with the normalization $\beta_A\alpha^A=1$ implying $FB-GA=1$ and choosing $F$ and $G$ such that in the Penrose limit the only non-vanishing component of the Weyl spinor is $\Psi(s)$.
Note $\alpha^A$ determines the flagpole, as well as together with  $\beta^A$ the flagplane \cite{PR1}.
Equally, as the transformation of the null tetrad vectors renders the only non vanishing Weyl scalar of the plane wave space time in the Penrose limit a real function \cite{chandra} this can also be achieved by transforming the spinor dyad $\{{\bf\beta},{\bf\alpha}\}$, namely, by $\beta^A\rightarrow e^{i\varphi_A}\beta^A$ and $\alpha^A\rightarrow e^{-i\varphi_A}\alpha^A$  \cite{PR1}.


\section{Example: Radial Penrose limit}
\label{sec4}
\setcounter{equation}{0}

The radial Penrose limit is an important particular case since the  affine parameter $s$ of the null geodesic becomes a function of just one of the null coordinates. 
In particular, the radial Penrose limit is taken along a null geodesic with tangent parallel to one of the real null tetrad vectors $l^{\mu}$ or $n^{\mu}$, respectively.
Here, the latter is chosen such that the null geodesic (\ref{nullgeo}) reads $V^{\mu}=|B|^2n^{\mu}$.
Equation (\ref{Bevol}) for $A\equiv 0$ and $B=|B|e^{i\varphi_B}$ with $\varphi_B$ a real function implies 
\begin{eqnarray}
\frac{1}{|B|}\frac{d|B|}{ds}&=&\frac{1}{4}\frac{dM}{ds}
\nonumber\\
\frac{d\varphi_B}{ds}&=&\frac{i}{4}\frac{1}{Z+\bar{Z}}\frac{d(Z-\bar{Z})}{ds}.
\end{eqnarray}
where equation (\ref{affinepar})(1) has been used.
Thus the Weyl scalar of the plane wave space-time is found to be 
\begin{eqnarray}
\Psi(s)=B^4(s)\Psi_4(s)=|B|^4|\Psi_4(s)|e^{i(4\varphi_B(s)+\varphi_{\Psi_4}(s))}
\end{eqnarray}
with 
\begin{eqnarray}
|B|&=&C_0e^{\frac{M}{4}}\\
e^{i\varphi_{\Psi_4}}&=&\left(\frac{1+i{\cal P}}{1-i{\cal P}}\right)^{\frac{1}{2}}\\
{\cal P}&=&-\frac{\frac{d^2\varphi_B}{ds^2}-\frac{1}{2}\frac{d}{ds}\left(2U+M+\ln(Z+\bar{Z})\right)
\frac{d\varphi_B}{ds}}{\frac{d^2}{ds^2}\ln(Z+\bar{Z})-\frac{1}{2}\frac{d}{ds}(2U+M)\frac{d}{ds}\ln(Z+\bar{Z})-8\left(\frac{d\varphi_B}{ds}\right)^2}.
\label{rad-sol}
\end{eqnarray}
and the affine parameter is determined by $s=C_0^{-2}\int du e^{-M(u,v=v_0)}$ with $C_0$ and $v_0$ constants. The plane wave metric in the radial Penrose limit is diagonal or diagonizable, respectively,
for $4\varphi_B+\varphi_{\Psi_4}$ equal to zero or constant, respectively. In the latter case the complex phase can be removed by a transformation of the spinor dyad as discussed in the previous section. As can be seen from the solution (\ref{rad-sol}) a constant phase $\varphi_B$ implies $\varphi_{\Psi_4}=0$. Moreover, it imposes $Z-\bar{Z}$ at most a constant implying the $G_2$ metric is diagonal (or diagonizable). Thus the radial Penrose limit is diagonal for diagonal $G_2$ space-times and not diagonal for non-diagonal ones. In \cite{kk03} an explicit example of this has been presented.


\section{Twistor equation and solutions in the Penrose limit}
\label{sec5}
\setcounter{equation}{0}

The twistor equation is given by \cite{PR2}
\begin{eqnarray}
\nabla_{A'}^{(A}\omega^{B)}=0.
\label{tw-eq}
\end{eqnarray}
Because of its conformal invariance solutions in Minkowski space-time can be transformed to solutions in conformally flat,  curved backgrounds.

In general in curved space-times solutions are severely restricted by the consistency condition
\begin{eqnarray}
\nabla^{A'(C}\nabla^A_{A'}\omega^{B)}=-\Box^{(CA}\omega^{B)}=-\Psi^{CA\hspace{0.25cm}B}_{\hspace{0.4cm} D}\omega^D-ie\varphi^{(CA}\omega^{B)},
\end{eqnarray}
permitting the presence of an electromagnetic field and a twistor $\omega^B$ with charge $e$.
Equation (\ref{tw-eq}) yields  the condition
\begin{eqnarray}
\Psi_{ABCD}\omega^D=-ie\varphi_{(AB}\omega_{C)}.
\label{cons-g2}
\end{eqnarray}
For uncharged twistors this implies that either the space-time is conformally flat ($\Psi_{ABCD}=0$) or
the Weyl spinor is null implying that it has a four-fold principal spinor. The latter is the case of plane wave space-times. 
Metrics of the form (\ref{metg2}) admit conformally flat  as  well as plane wave solutions. 
Writing the uncharged twistor $\omega^A= \omega^0o^A+\omega^1\iota^A$ its equation  is given by
\cite{PR1} in the GHP formalism
\begin{eqnarray}
\kappa\omega^0=\textup{\th}\omega^1, \hspace{1cm}
\sigma\omega^0&=&\textup{\dh}\omega^1,\hspace{1cm}\
\textup{\dh}'\omega^0=\sigma'\omega^1,  \hspace{1cm}\
\textup{\th}'\omega^0=\kappa'\omega^1,
\nonumber\\
\textup{\th}\omega^0+\rho\omega^0&=&\textup{\dh}'\omega^1+\tau'\omega^1,\hspace{2cm}\
\textup{\dh}\omega^0+\tau\omega^0=\textup{\th}'\omega^1+\rho'\omega^1.
\label{tw-ghp}
\end{eqnarray}
For the metric (\ref{metg2}) equation (\ref{tw-ghp}) yields to 
\begin{eqnarray}
\partial_u\omega^0&=&-\frac{1}{4}\left(\partial_uM-\frac{\partial_u(Z-\bar{Z})}{Z+\bar{Z}}
\right)\omega^0
\label{tw-g2-eq1}\\
\partial_v\omega^1&=&-\frac{1}{4}\left(\partial_vM+\frac{\partial_v(Z-\bar{Z})}{Z+\bar{Z}}
\right)\omega^1\\
\partial_x\omega^0&=&\frac{e^{\frac{1}{2}(M-U)}}{(Z+\bar{Z})^{\frac{1}{2}}}\left(\partial_u\omega^1-\frac{1}{4}\left(\partial_uM-2\partial_uU-\frac{\partial_u(Z-\bar{Z})}{Z+\bar{Z}}+4\frac{\partial_uZ}{Z+\bar{Z}}
\right)\omega^1\right)\\
-i\partial_y\omega^0&=&\frac{e^{\frac{1}{2}(M-U)}}{(Z+\bar{Z})^{\frac{1}{2}}}\left(Z
\partial_u\omega^1-\frac{Z}{4}\left(\partial_uM-2\partial_uU-\frac{\partial_u(Z-\bar{Z})}{Z+\bar{Z}}
\right)\omega^1
-\bar{Z}\frac{\partial_uZ}{Z+\bar{Z}}
\omega^1\right)\\
\partial_x\omega^1&=&\frac{e^{\frac{1}{2}(M-U)}}{(Z+\bar{Z})^{\frac{1}{2}}}\left(\partial_v\omega^0-\frac{1}{4}\left(\partial_vM-2\partial_vU+\frac{\partial_v(Z-\bar{Z})}{Z+\bar{Z}}+4\frac{\partial_v\bar{Z}}{Z+\bar{Z}}
\right)\omega^0\right)\\
i\partial_y\omega^1&=&\frac{e^{\frac{1}{2}(M-U)}}{(Z+\bar{Z})^{\frac{1}{2}}}\left(\bar{Z}
\partial_v\omega^0-\frac{\bar{Z}}{4}\left(\partial_vM-2\partial_vU+\frac{\partial_v(Z-\bar{Z})}{Z+\bar{Z}}
\right)\omega^0
-Z\frac{\partial_v\bar{Z}}{Z+\bar{Z}}
\omega^0\right)
\label{tw-g2-eq6}
\end{eqnarray}

The consistency condition for uncharged twistors $\omega^A$ in the $G_2$ background (cf. equation (\ref{cons-g2}) with $e\equiv 0$), yields in the Penrose limit $\Psi_{ABCD}\omega^D\alpha^A\alpha^B\alpha^C=0$ implying
\begin{eqnarray}
A(s)\omega^0\left(A^2(s)\Psi_0+B^2(s)\Psi_2\right)+B(s)\omega^1\left(A^2(s)\Psi_2+B^2(s)\Psi_4\right)=0.
\label{cons-pen-lim}
\end{eqnarray}
Assuming the space-time is not conformally flat equation (\ref{cons-pen-lim}) yields to  solutions 
$\omega^A$ of the twistor equation for
\begin{enumerate}
\item $A(s)\equiv 0$, $\omega^1\equiv 0$

Equations (\ref{tw-g2-eq1})-(\ref{tw-g2-eq6}) yield to
\begin{eqnarray}
\omega^0(u)=\exp\left(-\frac{1}{4}M(u)+\frac{1}{4}\int^{u}d\tilde{u}\frac{\partial_{\tilde{u}}
(Z-\bar{Z})}{Z+\bar{Z}}\right)
\end{eqnarray}
and $Z=Z(u)$, $M=M(u)$, $U=U(u)$ implying $\Psi_0\equiv 0$ and $\Psi_2\equiv 0$.

Taking into account the source-free Maxwell equations the solution is of the same general form. However, the solutions for the metric functions are different in general because of the  contribution from  the electromagnetic spinor component
$\varphi_2(u)=e^{\frac{1}{2}(M+U)}H_{,u}$ with $H=H(u)$
to equations  (\ref{max-e1})-(\ref{max-e4}).

\item $B(s)\equiv 0$, $\omega^0\equiv 0$. 
The solution is given by
\begin{eqnarray}
\omega^1(v)=\exp\left(-\frac{1}{4}M(v)-\frac{1}{4}\int^{v}d\tilde{v}\frac{\partial_{\tilde{v}}
(Z-\bar{Z})}{Z+\bar{Z}}\right)
\end{eqnarray}
and $Z=Z(v)$, $M=M(v)$, $U=U(v)$  implying $\Psi_2\equiv 0$ and $\Psi_4\equiv 0$.

Including a source-free electromagnetic field implies that the spinor component
$\varphi_0(v)=-e^{\frac{1}{2}(M+U)}H_{,v}$ with $H=H(v)$ contributes to equations 
(\ref{max-e1})-(\ref{max-e4}).

\end{enumerate}


\section{Conclusions}
\label{sec6}
\setcounter{equation}{0}

Penrose limits of $G_2$ space-times in vacuum as well as for source-free electromagnetic fields have been considered  using the formulation of Tod \cite{Tod2020} in the spinor formalism. Moreover, the condition for the diagonalizabilty of the resulting plane wave space-time has been considered for $G_2$ spacetimes. In terms of the Brinkmann form of the plane wave metric the non-diagonal wave profile is determined by the imaginary part of the only non-vanishing Weyl scalar. The Tod condition implies that the complex phase can be at most constant \cite{Tod2020}.
This could also be seen by arguing that a transformation of the null tetrad vectors renders the Weyl scalar to be a real function for a constant complex phase. As an example the radial Penrose limit has been considered in detail.
Finally, the twistor equation in the Penrose limit has been considered. Explicit solutions including a Maxwell field for uncharged twistors have been found in the radial Penrose limit. This points towards an additional, interesting aspect of Penrose limits. In general the consistency condition of the twistor equation severely restricts solutions in arbitrary, curved space-times. However, it is possible to associate corresponding twistor solutions in a Penrose limit of a general curved space-time.


\section{Acknowlegements}

Financial support by Spanish Science Ministry grant PID2021-123703NB-C22 (MCIU/AEI/FEDER, EU) and Basque Government grant IT1628-22 is gratefully acknowledged.


\appendix
\section{Quantities in the Newman-Penrose formalism}
\label{appA}
\renewcommand{\theequation}{\thesection.\arabic{equation}}

The relevant quantities for the metric ({\ref{metg2}) are given here some of which have been calculated using the {\tt mathematica} package {\tt xAct} \cite{xAct}.
The null tetrad metric is given by
\begin{eqnarray}
\eta_{(a)(b)}=\eta^{(a)(b)}=\left(
\begin{array}{c c  r r}
 0 & 1 &  0 &  0\\
 1 & 0 &   0 &  0\\
 0 & 0 &   0 & -1\\
 0 & 0  & -1 & 0
\end{array}
\right)
\end{eqnarray}
Tetrad indices are latin indices running from 1 to 4, enclosed in brackets.
The null tetrad vectors for the metric (\ref{metg2}) are given by
\begin{eqnarray}
l^{\mu}=\left(
0\;\; e^{\frac{M}{2}}\;\;0\;\;0\right)
&\hspace{1.6cm}&
n^{\mu}=\left(
 e^{\frac{M}{2}}\;\;0\;\;0\;\;0\right)
\nonumber\\
m^{\mu}=\left(
0\;\;0\;\; \frac{e^{\frac{U}{2}}\bar{Z}}{\left(Z+\bar{Z}\right)^{\frac{1}{2}}}
\;\;
-i\frac{e^{\frac{U}{2}}}{\left(Z+\bar{Z}\right)^{\frac{1}{2}}}
\right)
&\hspace{0.6cm}&
\bar{m}^{\mu}=\left(
0\;\;0\;\; \frac{e^{\frac{U}{2}}Z}{\left(Z+\bar{Z}\right)^{\frac{1}{2}}}
\;\;
i\frac{e^{\frac{U}{2}}}{\left(Z+\bar{Z}\right)^{\frac{1}{2}}}
\right)
\end{eqnarray}
Using the notation of \cite{PR1} the directional derivatives are defined by 
\begin{eqnarray}
D = l^{\nu}\nabla_{\nu}\hspace{1.6cm}
D' = n^{\nu}\nabla_{\nu}\hspace{1.6cm}
\delta = m^{\nu}\nabla_{\nu}\hspace{1.6cm}
\delta' = \bar{m}^{\nu}\nabla_{\nu} .
\end{eqnarray}
For the definitions of the spin-weighted directional derivatives as used in the compacted spin-coefficient (GHP) formalism, $\textup{\th}$, $\textup{\th}'$, $\textup{\dh}$ and $\textup{\dh}'$, cf. \cite{PR1} and references therein.

The only non vanishing spin coefficients are 
\begin{eqnarray}
\begin{array}{l l }
\epsilon=-\frac{1}{4}e^{\frac{1}{2}M(u,v)}\left[\partial_vM+\frac{\partial_v\left(Z-\bar{Z}\right)}
{Z+\bar{Z}}\right] &\hspace{1cm} 
\gamma=\frac{1}{4}e^{\frac{1}{2}M(u,v)}\left[\partial_u M-\frac{\partial_u\left(Z-\bar{Z}\right)}
{Z+\bar{Z}}\right]
\\
\rho'=\frac{1}{2}e^{\frac{1}{2}M(u,v)}\partial_uU &\hspace{1cm} 
\sigma'=e^{\frac{1}{2}M(u,v)}\frac{\partial_uZ}{Z+\bar{Z}}
\\
\rho=\frac{1}{2}e^{\frac{1}{2}M(u,v)}\partial_vU  &\hspace{1cm} 
\sigma=e^{\frac{1}{2}M(u,v)}\frac{\partial_v\bar{Z}}{Z+\bar{Z}}
\end{array}
\label{spincoef}
\end{eqnarray}
The components of the Weyl tensor are encoded in 5 complex Weyl scalars in the 
Newmann-Penrose formalism. The non vanishing Weyl scalars for the metric $G_2$ metric 
(\ref{metg2}) are given by
\begin{eqnarray}
\Psi_0&=&\frac{e^M}{(Z+\bar{Z})^2}\left[
(Z+\bar{Z})[\bar{Z}_{,vv}+M_{,v}\bar{Z}_{,v}-U_{,v}\bar{Z}_{,v}]-2(\bar{Z}_{,v})^2
\right]
\label{psi0}\\
\Psi_2&=&-\frac{e^M}{4}\left[2U_{,uv}-U_{,u}U_{,v}-4\frac{Z_{,u}\bar{Z}_{,v}}{(Z+\bar{Z})^2}
\right]-2\Pi
\label{psi2}
\\
\Psi_4&=&\frac{e^M}{(Z+\bar{Z})^2}\left[
(Z+\bar{Z})[Z_{,uu}+M_{,u}Z_{,u}-U_{,u}Z_{,u}]-2(Z_{,u})^2
\right]
\label{psi4}
\end{eqnarray}
with the notation $X_{,m}\equiv\frac{\partial X}{\partial m}$ and 
$X_{,mn} \equiv\frac{\partial^2 X}{\partial m\partial n}$ for $m,n$ denoting the null variables $u$ and $v$.
Moreover, in the expression for the Weyl scalar $\Psi_2$ (\ref{psi2}) $\Pi=\Lambda=\frac{1}{24}R$ with  $R$ the Ricci scalar, as given below.

The components of the Ricci tensor are encoded in the 4 real and 3 complex scalars given of which the following are non-vanishing for the $G_2$ metric under consideration,
\begin{eqnarray}
\Phi_{00}&=&\frac{e^M}{4}\left[2U_{,vv}-U^2_{,v}+2M_{,v}U_{,v}-4\frac{Z_{,v}\bar{Z}_{,v}}{(Z+\bar{Z})^2}\right]
\label{phi00}\\
\Phi_{02}&=&-\frac{1}{2}\frac{e^M}{Z+\bar{Z}}\left[2\bar{Z}_{,uv}-\bar{Z}_{,u}U_{,v}-\bar{Z}_{,v}U_{,u}
-4\frac{\bar{Z}_{,u}\bar{Z}_{,v}}{Z+\bar{Z}}\right]\\
\Phi_{20}&=&\bar{\Phi}_{02}\\
\Phi_{11}&=&\frac{e^M}{8}\left[U_{,u}U_{,v}+2M_{,uv}-2\frac{Z_{,u}\bar{Z}_{,v}+\bar{Z}_{,u}Z_{,v}}
{(Z+\bar{Z})^2}\right]\\
\Phi_{22}&=&\frac{e^M}{4}\left[2U_{,uu}-U_{,u}^2+2M_{,u}U_{,u}-4\frac{Z_{,u}\bar{Z}_{,u}}{(Z+\bar{Z})^2}
\right]
\label{phi22}\\
\Lambda&=&-\frac{e^M}{24}\left[4U_{,uv}-3U_{,u}U_{,v}+2M_{,uv}-2\frac{Z_{,u}\bar{Z}_{,v}+\bar{Z}_{,u}Z_{,v}}
{(Z+\bar{Z})^2}\right].
\end{eqnarray}


\bibliography{references}

\bibliographystyle{apsrev}

\end{document}